\documentclass[]{article}
\usepackage{emulateapj,onecolfloat,psfig}

\newcommand{\eg}{{\it e.g.}}
\newcommand{\etal}{{\it et al.}}
\newcommand{\etc}{{\it etc.}}
\newcommand{\ie}{{\it i.e.}}
\newcommand{\DF}{{\sc DF}}

\newcommand{\kms}{\ensuremath{\mathrm{km}\,\mathrm{s}^{-1}}}
\newcommand{\LCDM}{$\Lambda$CDM}
\newcommand{\ML}{\ensuremath{\Upsilon_{*}}}
\newcommand{\MLsun}{\ensuremath{{M}_{\odot}/L_{\odot}}}
\newcommand{\Q}{\ensuremath{{\cal Q}}}
\newcommand{\Vf}{\ensuremath{V_{f}}}
\newcommand{\ld}{\ensuremath{\ell_{d}}}
\newcommand{\md}{\ensuremath{m_{d}}}
\newcommand{\rd}{\ensuremath{R_{d}}}
\newcommand{\rs}{\ensuremath{r_{s}}}
\newcommand{\VNFW}{\ensuremath{V_{200}}}

\slugcomment{\it Revised version submitted to {\it The Astrophysical Journal}}

\begin{document}

\twocolumn
[
\title{The Compression of Dark Matter Halos by Baryonic Infall}
\author{J. A. Sellwood}
\affil{Rutgers University, Department of Physics \& Astronomy, \\
       136 Frelinghuysen Road, Piscataway, NJ 08854-8019 \\
       {\it sellwood@physics.rutgers.edu}}
\and
\author{Stacy S. McGaugh}
\affil{Department of Astronomy, University of Maryland, \ 
       College Park, MD 20742-2421 \\
       {\it ssm@astro.umd.edu}}

\begin{abstract}
The initial radial density profiles of dark matter halos are laid down
by gravitational collapse in hierarchical structure formation
scenarios and are subject to further compression as baryons cool and
settle to the halo centers.  We here describe an explicit
implementation of the algorithm, originally developed by Young, to
calculate changes to the density profile as the result of adiabatic
infall in a spherical halo model.  Halos with random motion are more
resistant to compression than are those in which random motions are
neglected, which is a key weakness of the simple method widely
employed.  Young's algorithm results in density profiles in excellent
agreement with those from $N$-body simulations.  We show how the
algorithm may be applied to determine the original uncompressed halos
of real galaxies, a step which must be computed with care in order to
enable a confrontation with theoretical predictions from theories such
as \LCDM.
\end{abstract}

 ]

\section{Introduction}
Initial density fluctuations in the early universe are widely believed
to seed the collapse of dark matter halos through gravitational
instability.  Computation of this process has allowed the clustering
and properties of the collapsed halos to be predicted in detail in
simulations of the collisionless component only (\eg, Navarro, Frenk
\& White 1997; Jing 2000; Bullock \etal\ 2001; Diemand \etal\ 2004;
Navarro \etal\ 2004).

Initially, the dark matter and hot gas are well-mixed (Spergel \etal\
2003), but galaxies are believed to form as the baryons cool and
settle towards the centers of collapsed halos, as originally envisaged
by White \& Rees (1978), Fall \& Efstathiou (1980), Gunn (1982), and
others.  The settling of gas towards the center causes further
compression of the halo.  This latter process is followed directly in
simulations that include gas cooling (Gottl\"ober \etal\ 2002; Abadi
\etal\ 2003; Governato \etal\ 2004), but an approximate analytic
treatment is needed for many applications.  The conventional procedure
(Blumenthal \etal\ 1986) is widely used to compute, for example, the
expected surface brightness (\eg, Dalcanton, Spergel \& Summers 1997;
Mo, Mao \& White 1998), more-elaborate semi-analytic galaxy formation
models (\eg, Cole \etal\ 2000; van den Bosch \& Dalcanton 2000),
predictions for lensing (\eg, Keeton 2001), predictions from tilted
power spectra (\eg, Zenter \& Bullock 2002), \etc\ \ It has been
reported that the Blumenthal formula generally over-estimates the
actual compression (Barnes 1987; Sellwood 1999; Gnedin \etal\ 2004).

However, Young (1980) showed that adiabatic compression of a spherical
system can be treated exactly, without having to run expensive
simulations.  His formulation was originally to model the growth of a
black hole in a spherical star cluster, but the method can be applied
for any adiabatic change to a spherical potential.  It was first used
for halo compression by Wilson (2004) and developed independently by
us.

We review the standard method for halo compression in
\S\ref{previous}, and describe Young's (1980) method in \S\ref{Young}.
We give a few examples and tests in \S\ref{tests} and illustrate the
application of the method to data from real galaxies in
\S\ref{applic}.

\section{Halo compression}
\label{previous}
The usual algorithm for halo compression is easy to implement, but
very crude.  It is generally attributed to Blumenthal \etal\ (1986),
although the same idea was developed independently by Barnes \& White
(1984), Ryden \& Gunn (1987), and others.  It assumes that dark matter
particles conserve only their angular momenta as the halo is
compressed, which is equivalent to assuming that all the halo
particles move on circular orbits (see below).  In a spherical
potential, the squared angular momentum of a circular orbit is $L^2 =
rGM(r)$, and if this quantity is conserved as a disk with mass profile
$M_d(r)$ grows slowly, we have
\begin{equation}
r_iM_i(r_i) = r_f[ M_d(r_f) + (1-f_d)M_f(r_f) ],
\label{blum1}
\end{equation}
where $M_i$ is the initial total mass (dark plus baryonic) profile,
$(1-f_d)M_f$ is the desired final dark matter mass profile, and $r_f$
is the final radius of the mass shell initially at radius $r_i$.  The
quantity $f_d$ is the fraction of the initial total mass, assumed to
be independent of radius, that condenses to form the disk.  We can
substitute for $M_f(r_f)$ by making use of the assumption
\begin{equation}
M_i(r_i) = M_f(r_f)
\label{blum2}
\end{equation}
which is sometimes stated that ``shells of matter do not cross''.  We
can then find $r_i$ for any desired $r_f$ and, through
eq.~(\ref{blum2}), we can obtain the mass profile of the compressed
dark matter halo.  For convenience, we denote this the Blumenthal
algorithm.

Eq. (\ref{blum1}) assumes that the disk mass is taken from the halo in
equal proportions at all radii.  A possible alternative assumption is
to imagine that the mass of the disk arrives from an external source,
in which case the right-hand side reads $r_f[ M_d(r_f) + M_f(r_f) ]$.
Scaling the compressed mass profile by the factor $1/(1+f_d)$ would
conserve total mass once more and results in a scaled density profile
that is not mathematically identical to that resulting from the more
common procedure, but differences are small everywhere.  Neither
hypothesis is likely to be correct, since the condensing mass fraction
will depend on factors such as the cooling rate at each radius, but
such refinements are unlikely to affect the inner halo density much.

Flores \etal\ (1993) report some $N$-body tests which indicated that
the Blumenthal algorithm yields a reasonable approximation to the
compressed halo, and Jesseit \etal\ (2002) reached a similar
conclusion.  But Barnes (1987), Sellwood (1999) and Gnedin \etal\
(2004) have warned that the predicted density profile is more
concentrated than found in their $N$-body simulations, especially in
the crucial inner part.  Gnedin \etal\ suggest that the above scheme
be amended to take account of the radial motions of the particles, and
recommend the substitution $M(r)r \rightarrow M(\bar r)r$ in
eq.~(\ref{blum2}), where $\bar r$ is the time-averaged radius of each
halo particle.  Their revised formula gives better agreement with
their simulations, in which halos are not perfectly relaxed, but
reduces to the Blumenthal prescription for equilibrium distribution
functions of spherical halos.

A number of authors have reported difficulties when trying to apply
the Blumenthal method to compute an original halo, from the observed
compressed form (\eg\ Weiner, Sellwood \& Williams 2001).  As long as
the adiabatic assumption holds, valid formulae for compression can be
employed to deduce the decompressed density.  However, Weiner \etal\
found that the Blumenthal algorithm fails for decompression because
the radial density profile can become multi-valued, which is both
physically impossible and inconsistent with the central assumption
that ``mass shells do not cross.''  Valid formulae for adiabatic
changes must apply irrespective of the direction in which the change
occurs, and the failure of the Blumenthal scheme in this regard is
further evidence of its inadequacy.

\section{Young's Method}
\label{Young}
As stressed in \S3.6 of Binney \& Tremaine (1987, hereafter BT),
adiabatic changes conserve all three actions of an orbit.  In a
spherical system, the actions are the angular momentum and radial
action, while the third action is identically zero because the plane
of each orbit is an invariant.  Thus adiabatic compression of a
spherical halo needs to take account of the conservation of radial
action, as well as of angular momentum.  (By ignoring the radial
action, the Blumenthal algorithm implicitly assumes it is zero, and
all particles move on circular orbits -- as is well known.)  Young
(1980) described an algorithm that takes proper account of the effects
of random motion in the halo, while still assuming spherical symmetry,
and similar considerations underlie the Fokker-Planck calculations of
globular cluster evolution (\eg, Cohn 1979).  Wilson (2004) has
already applied the scheme we outline here for halo compression.

The assumption of spherical symmetry is crucial, since without it we
would need to consider three actions, which would make the problem
essentially intractable analytically -- although it could still be
followed in $N$-body simulations.  Jesseit \etal\ (2002) and Wilson
(2004) tested this assumption using $N$-body simulations, and we
report an additional test in \S\ref{tests} that indeed shows that the
spherically averaged density profile agrees extremely well at all
radii with the predicted change from a Young-type code.  The
effectiveness of the spherical assumption results from the fact that
even quite heavy disks lead to very mild flattening of an isotropic
halo, which justifies neglect of the third action.

The central idea of Young's method is to apply the constraint that the
value of the distribution function (\DF) expressed as a function of
the actions is invariant during adiabatic changes -- \ie\
$f_0(J_r,J_\phi) = f_n(J_r,J_\phi)$.  In this formula, $J_r(E,L)$ is
the radial action, and $J_\phi \equiv L$ is the azimuthal action, or
total angular momentum per unit mass, and the subscripts 0 \& $n$
refer respectively to the original and new halo profiles.  While $f$
expressed as a function of the actions does not change, $f_n(E_n,L)$
does change because the relation between the radial action, $J_r$, and
the specific energy, $E_n$, depends on the potential well.

Starting from some spherical initial model with density, $\rho_0(r)$,
potential $\Phi_0(r)$, and \DF\ $f_0(E_0,L)$, we add a second mass
component, and search iteratively for new functions $\rho_n(r)$,
$\Phi_n(r)$, and $f_n(E_n,L)$.  At the $n$-th iteration, we determine
a new spherical density profile from
\begin{equation}
\rho_{n+1}(r) = 4\pi \int_{\Phi_n(r)}^{\Phi_n(\infty)} \int_0^{L_{\rm
max}} {L f_n(E_n,L) \over r^2 u}
\; dL dE_n,
\label{renew_rho}
\end{equation}
where the radial speed
\begin{equation}
u = \{2[E_n-\Phi_n(r)] - L^2 / r^2\}^{1/2},
\end{equation}
and $L_{\rm max}(E) = r\{2[E-\Phi_n(r)]\}^{1/2}$.  Since the value of
$f_n$ to be used in the integrand is determined by the condition
$f_n(J_r,J_\phi) = f_0(J_r,J_\phi)$, we need to know $J_r(E_n,L)$ in
the $n$-th potential well, and then to look up $E_0(J_r,L)$ in the
initial potential well $\Phi_0(r)$, in order to evaluate $f_0(E_0,L)$.
As these functions are not available in closed form, except for
special potentials such as the isochrone (Eggen, Lynden-Bell \&
Sandage 1962), we determine both by interpolation in 2-D tables of
values; the table for $J_r(E_n,L)$ has to be updated at every
iteration.

Since $\rho_{n+1}(r)$ is spherically symmetric, we have for halo
compression
\begin{equation}
\Phi_{n+1}(r) = -G \int_r^\infty {M_{n+1}(r^\prime) \over r^{\prime2}}
dr^\prime + \Phi_{\rm ext}(r).
\label{renew_phi}
\end{equation}
Note that $\Phi_{n+1}(r)$ includes $\Phi_{\rm ext}$, the monopole term
only arising from the disk mass distribution.  It is most convenient
to create, and keep updated, 1-D tables for both $\Phi_n(r)$ and
$\rho_n(r)$.

This new estimate of the gravitational potential is then used in
eq.~(\ref{renew_rho}) to obtain an improved estimate of the halo
density, and the whole procedure is iterated until the solution
converges.  In practice, between ten and twenty iterations are needed
for the potential profile to stabilize to a part in $10^5$.  The
method conserves mass, the final mass of compressed halo differs from
the initial by a few parts in $10^4$ for the grids we typically
employ.

The disk mass can easily be subtracted from the halo if desired.  All
that is needed is to normalize the density given by
eq.~(\ref{renew_rho}) by the factor $(M_h - M_d)/M_h$.  Here, $M_h$ is
the mass of the halo to the virial radius.  Alternatively, as we
described above for the Blumenthal algorithm, one can re-normalize the
combined disk and halo masses to equal the original total mass.

This algorithm is clearly more difficult to program, and more
time-consuming to run, than the Blumenthal algorithm.  We typically
use tables for $\rho_n$, $\Phi_n$, \etc, that have 100 radial points.
The table for $J_r(E_n,L)$, which has to be rebuilt at every
iteration, has $100 \times 50$ values, which when fitted with a
bicubic spline achieves interpolated estimates good to a part in
$10^7$ or better; constructing this table takes the lion share of the
computational effort.  Nevertheless, the whole iterative scheme
converges within a few cpu minutes.

Young (1980) shows that a given mass profile is less compressed when
an isotropic \DF\ is assumed than when all particles have circular
orbits, which is the implicit assumption of the Blumenthal algorithm.
Thus allowance for radial motion makes a dark matter halo somewhat
more resistant to compression.

It should also be noted that even if the initial \DF\ were isotropic,
the compressed \DF\ will not be.  Since the \DF\ of the compressed
halo is computed anyway, Young's algorithm yields direct estimates of
the radial anisotropy expected.

\begin{figure}[t]
\centerline{\psfig{figure=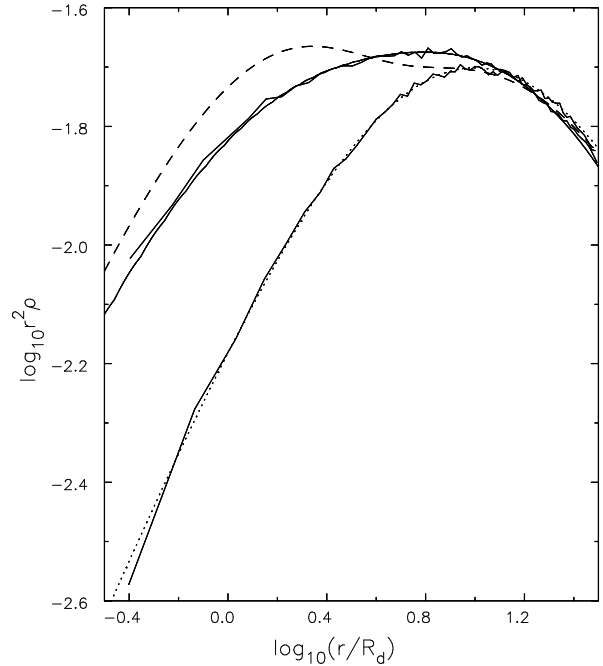,width=.9\hsize,clip=}}
\caption{Test of the spherical approximation for a disk grown in an
NFW halo.  The jagged lines show the density profile estimated from
the particles before and after a thin disk is grown.  The dotted and
smooth solid lines show respectively the density of the analytic NFW
halo and the compressed profile computed from Young's algorithm.  The
dashed curve shows the compression predicted by the Blumenthal
algorithm, which is clearly inconsistent with the $N$-body results.}
\label{sphcheck}
\end{figure}

\begin{figure*}[t]
\centerline{\psfig{figure=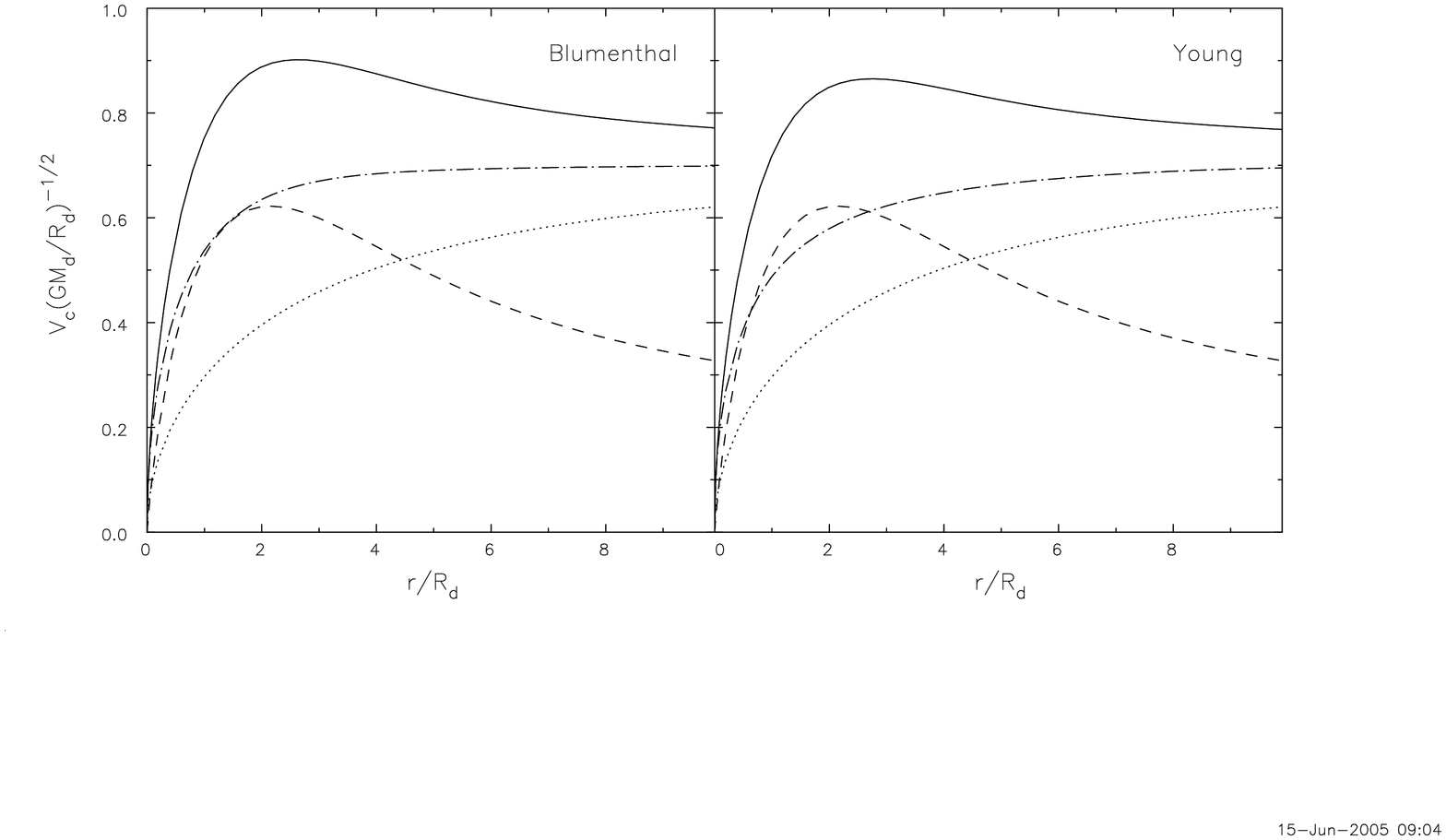,width=.9\hsize,angle=270,clip=}}
\caption{Comparison between halo compression using the Blumenthal
algorithm on the left, and Young's formula on the right, showing that
halos are less easily compressed when velocity dispersion is taken
into account.  In both panels, the solid curve shows the total
circular speed, the dashed curve is the disk contribution, the dotted
curve is the uncompressed NFW halo, and the dot-dash curve is the
compressed halo contribution.  Note that $\md = 0.05$ in this case.}
\label{contrast}
\end{figure*}

The method can, in principle, be adapted for halo decompression.
Provided an equilibrium \DF\ for the compressed halo can be
constructed, the algorithm guarantees that the decompressed halo must
have a positive, single-valued, density everywhere.  The only
significant barrier to extending the method to this application is the
construction from observational data of suitable \DF s for compressed
halos containing disks.

Young's method is not restricted to any particular radial mass
profiles for either the disk or the halo; the added external mass can
have any spherically averaged mass profile.  Neither is the method
restricted to isotropic \DF s; any $f(E,L)$ that yields the initial
spherical mass profile may be used.

\section{Examples and tests}
\label{tests}
Figure \ref{sphcheck} presents results from an $N$-body simulation
designed to test the adequacy of the spherical approximation for the
disk potential.  This test uses the NFW density profile (Navarro,
Frenk \& White 1997)
\begin{equation}
\rho_N(r) = {\rho_s \rs^3 \over r(r+\rs)^2},
\label{rhoNFW}
\end{equation}
with $\rho_s$ setting the density scale at the break radius, $\rs$.
We define a mass unit as $M_N = 4\pi \rho_s \rs^3$, which is equal to
the mass enclosed within $r \simeq 5.3054\rs$.  The radial mass
profile of an exponential disk is
\begin{equation}
m(R) = M_d \left[ 1 - \left( 1 + {R \over \rd} \right) \exp\left(-{R
\over \rd} \right) \right],
\end{equation}
where $M_d$ is the total mass of the disk and $\rd$ is the disk scale
length.  We characterize a simple disk-halo model by two parameters:
the mass ratio $\md \equiv M_d/M_N$ and length ratio $\ld \equiv
\rd/\rs$.

The isotropic \DF\ for an NFW halo is given by Eddington's inversion
formula (BT, \S4.4), and is positive everywhere.  This equation has to
be solved numerically for the NFW potential, which requires a
sophisticated quadrature method.  It is convenient to impose an upper
energy cut-off, so that the halo density declines to zero at a large
radius, but the boundary term should not be included.  Alternatively,
one can use Widrow's (2000) fitted approximation for the \DF.

The halo is represented by 1 million particles selected smoothly from
the \DF\ as described in Appendix B of Debattista \& Sellwood (2000);
in order to reduce shot noise when estimating the density in the inner
halo, the halo particle masses were scaled as $\sqrt{L}$, where $L$ is
the total angular momentum of the particle.  The disk in the
simulation is represented by 100\,000 equal mass particles that are
added at a constant rate in the $z=0$ plane and held fixed in their
initial positions in order that the disk mass profile remains exactly
exponential.  The halo particles move in response to forces from both
the disk and halo, which are computed on a spherical grid of 500
radial shells (McGlynn 1984; Sellwood 2003, Appendix A) and include
all multipole terms $0 \leq l \leq 8$.  The disk is grown over the
period $0 \leq t \leq 250(R_d^3/GM_d)$ and the final density shown in
Fig.~\ref{sphcheck} is measured after twice this period.  The time
step was $0.005(R_d^3/GM_d)$.

The dotted curve in Fig.~\ref{sphcheck} shows the density function
(eq.~\ref{rhoNFW}) that is closely followed by the initial density
profile of the 1M particles (jagged line).  The smooth solid line
shows the predicted density profile from Young's algorithm after an
exponential disk with $\ld=0.1$ and $\md = 0.05$ is grown slowly.  The
jagged solid line that follows this curve closely is the (spherically
averaged) density profile measured from the particles when the
asphericity of the thin disk and halo are taken into account.
Evidently, the spherical approximation is quite adequate in Young's
algorithm.

The dashed curve in Fig.~\ref{sphcheck} shows the prediction from the
Blumenthal algorithm, which is clearly inconsistent with the $N$-body
results.  This discrepancy confirms the conclusion reached previously
by Barnes (1987), Sellwood (1999) and Gnedin \etal\ (2004) that the
Blumenthal algorithm over-estimates halo compression in the inner
parts.

A more conventional indication of the difference between the
predictions by Young's and Blumenthal's methods is illustrated in
Figure \ref{contrast} for the same model shown in Fig.~\ref{sphcheck}.
The compressed halo density (dot-dashed curve) predicted by the
Blumenthal algorithm is shown on the left, while the right-hand panel
shows the prediction from Young's procedure.  Note that the greatest
differences occur in the inner region where the disk contributes,
which is generally the region most strongly constrained by observed
rotation curves.

\begin{figure}[t]
\centerline{\psfig{figure=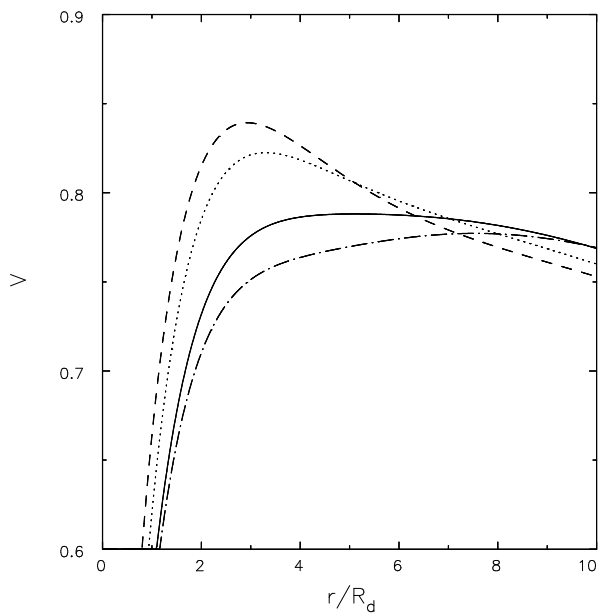,width=.9\hsize,clip=}}
\caption{The effect of velocity anisotropy in a Plummer halo.  The
dashed curve shows the circular speed resulting from compression using
the Blumenthal algorithm while the other curves show the same quantity
computed by Young's method for different initial DFs.  The solid curve
is for an isotropic model, the dot-dashed curve for a model with
maximum radial bias, and the dotted curve shows a strongly azimuthally
biased model.  Note the suppressed zero on the velocity axis -- all
curves continue approximately linearly to the origin.}
\label{anisot}
\end{figure}

Figure \ref{anisot} shows the effect of changing the velocity
distribution in a Plummer halo model.  In all cases shown, the Plummer
sphere is compressed adiabatically by addition of an exponential disk
having one tenth the mass of the halo, and disk scale length one tenth
of the Plummer core radius.  Dejonghe (1987) provides a family of \DF
s for this model with a single parameter, $q$, that determines the
shape of the velocity distribution.  The solid curve shows the
resulting rotation curve when the halo has an isotropic \DF\ ($q=0$).
The dot-dashed curve shows that a \DF\ that is maximally radially
biased ($q=2$) is compressed somewhat less, while the dotted curve
shows the result when a strongly azimuthally biased \DF\ ($q=-15$) is
assumed.  The dashed curve shows the result when the Blumenthal
algorithm is employed, again showing that its assumption of extreme
azimuthal bias in the orbits of halo particles allows too strong a
compression of the halo.

The compressed density resulting from Young's and Blumenthal's methods
generally shows larger differences for models with cores, such as King
and Plummer models, than for cusped density profiles.  However, even
for the NFW profile (Fig.~\ref{contrast}), the differences remain
quite significant.

\section{Application of Young's Method to Data}
\label{applic}
In this section, we use Young's algorithm to deduce the initial halos
of five galaxies from their observed rotation curves.  We present
these results here purely to illustrate an application to real data,
and will report a more systematic study in a later paper.

We selected galaxies with extended HI rotation curves from the
compilation of Sanders \& McGaugh (2002) so that the shape of the
rotation curve is measured well into the halo.  In order to explore a
range of galaxy properties, we selected five galaxies that span a
large range in both luminosity and surface brightness.  The five
objects are the bright galaxy NGC 2903, the intermediate luminosity
high and low surface brightness galaxies NGC 2403 and UGC 128, and the
high and low surface brightness dwarf galaxies NGC 55 and NGC 1560.

\begin{deluxetable}{lcccccccccr}
\tablewidth{0pt}
\tablecaption{Disk, Compressed, and Primordial Halo Parameters\label{datafits}}
\tablehead{
\colhead{Galaxy} & \colhead{\Vf\tablenotemark{a}} & \colhead{\rd\tablenotemark{b}} &
  \colhead{\ML\tablenotemark{c}} & \colhead{${\cal Q}$} & 
  \colhead{\md} & \colhead{\ld} & \colhead{\rs\tablenotemark{b}} & 
  \colhead{$c$} & \colhead{\VNFW\tablenotemark{a}} & \colhead{Refs.} }
%
%
\startdata
NGC 2903 & 185 & \phn2.0 & 2.2 &  0.6 & 0.10 & 0.12  & 16.7  
 & \phn9.2 & 115 & 1,2 \\
NGC 2403 & 134 & \phn2.1 & 0.6 & 0.4 & 0.02 & 0.15 & 14.0 
 & 10.1 & 106 & 1,2,3 \\
UGC \phn128 & 131 & 10\tablenotemark{d}\phd\phn & 1.1 & 1.0 & 0.02
 & 0.19 & 52.6 & \phn3.2 & 126 & 4,5 \\
NGC \phn\phn55 & \phn86 & \phn2.7\tablenotemark{d} & 0.2 & 1.0 & 0.01
 & 0.10 & 27.0 & \phn4.6 & \phn93 & 6 \\ 
NGC 1560 & \phn72 & \phn2.6\tablenotemark{d} & 1.0 & 1.0 & 0.01 
 & 0.12 & 21.7 & \phn4.9  &  \phn80 & 7 \\
\enddata
\tablenotetext{a}{In \kms.}
\tablenotetext{b}{In kpc.}
\tablenotetext{c}{In solar units, \MLsun,  in the $B$-band.}
\tablenotetext{d}{This baryonic scale length is rather larger than that
 of the stars alone.}
\tablerefs{1. Begeman (1987).
2. Wevers, van der Kruit \& Allen (1986).
3. Blais-Ouellette \etal\ (2004).
4. van der Hulst \etal\ (1993).
5. de Blok, van der Hulst \& Bothun (1995).
6. Puche, Carignan, Wainscoat (1991).
7. Broeils (1992).}
\end{deluxetable}

We stress that these galaxies were selected to be purely illustrative
of our technique over a wide range of galaxy properties, and do not
constitute a representative sample.  They do, however, provide some
interesting insights.

We assume an exponential disk, which is adequate for all five galaxies
in this very preliminary study, and an NFW halo with an isotropic \DF.
We have to choose values for our two parameters: the length and mass
ratios, \ld\ and \md, of the disk and halo.  The model halo is
dimensionless, so the physical parameters of the primordial halo can
be inferred only after scaling to fit a particular galaxy.  The
conversion to ($c$, \VNFW) depends weakly on the Hubble constant; we
adopt $H_0 = 75\,\textrm{km}\,\textrm{s}^{-1}\,\textrm{Mpc}^{-1}$ for
consistency with the galaxy data.

We first constructed a grid of compressed NFW models with a range of
plausible \md\ and \ld\ and examined the final rotation curves from
this grid to judge which combinations of disk parameters might make
tolerable approximations to the data.  Crudely speaking, $\md \le 0.1$
and $\ld \le 0.2$ in NFW halos yield plausible-looking rotation
curves.  Larger \md\ compresses the halo too much, leading to rotation
curves that have a pronounced peak and fall too far before leveling
off.  Larger \ld\ causes the disk to extend too far out into the halo
where the declining portion of the rotation curve is too obvious
compared to data.  This is only an approximate statement: some
individual galaxies might be described by more extreme parameters, and
different halo types could well require very different combinations of
\md\ and \ld.

A trend in our grid of models mimics one that is well-established by
the data.  The rotation curves rise quickly and fall gradually for
larger values of \md, qualitatively as observed for luminous spirals.
The disk-driven peak declines as we employ progressively lower \md,
qualitatively giving the gradual, and continuing, rise characteristic
of dwarf galaxy rotation curves.  Varying \ld\ at fixed \md\ adjusts
the amplitude of the total rotation curve relative to that of the
baryons.

We used this grid to estimate plausible parameters for a first
approximation to the data for each case, and then refine the model
iteratively.  At this juncture, goodness of fit is judged by eye.  Our
models match the data reasonably well, though are certainly not
perfect.  The shortcomings of the fits can in some cases be attributed
to deviations from our idealized exponential disk model, which we
assume here for simplicity.  Indeed, we expect to improve the fits by
using the exact disk mass profile, but leave this refinement to future
work, as the dominant uncertainty is not in the precise shape of the
baryonic mass distribution but in its amplitude, which depends on the
stellar mass-to-light ratio \ML.  For a specific choice of \ML,
matching both the total and baryonic rotation curves simultaneously
restricts the plausible values of \md\ and \ld\ fairly well, though we
make no claim that our fits are unique.

We must adopt some estimate for the baryonic mass in each galaxy,
which is straightforward for the gas, but fraught for stars.  McGaugh
(2004) discusses the \ML\ for all the galaxies considered here,
preferring a prescription that minimizes the scatter both in the
baryonic Tully-Fisher relation (McGaugh \etal\ 2000; McGaugh 2005) and
in the mass discrepancy-acceleration relation.  He defines the ratio
of the adopted \ML\ to this optimal one is given by the parameter $\Q
= \ML/\Upsilon_{opt}$, which encapsulates the goodness of the
mass-to-light ratio in this respect: the further \Q\ is removed from
unity, the larger the scatter induced in these otherwise tight
correlations.

The basic galaxy data and results are given in Table \ref{datafits}.
The first column gives each galaxy's name.  The second column gives
the observed flat rotation velocity.  The third column gives the
physical scale length of the exponential disk chosen to approximate
each galaxy's baryon profile; since it is the sum of stars and gas
that matters, the scale length has been stretched where necessary to
approximate the total.  Column 4 gives our final adopted mass-to-light
ratio \ML\ (in the $B$-band) for the stars.  Where possible, this has
been held fixed at the optimal value described by McGaugh (2004), but
in some cases we had to reduce \ML\ in order to obtain a fit, and we
give the deviation from the optimal value in column 5.  This deviation
is inevitably in the sense that the disk mass must be reduced in order
to accommodate the cusp of the NFW halo.  Columns 6 and 7 give the
halo parameters \md\ and \ld\ of our adopted fit.  The scale length of
the halo \rs\ corresponding to the baryonic scale length \rd\ and
fitted \ld\ is given in column 8.  The NFW concentration and \VNFW\
parameters of the primordial halo are given in columns 9 and 10.
References to the original sources of the data used are given in
column 11.

\begin{figure}[t]
\centerline{\psfig{figure=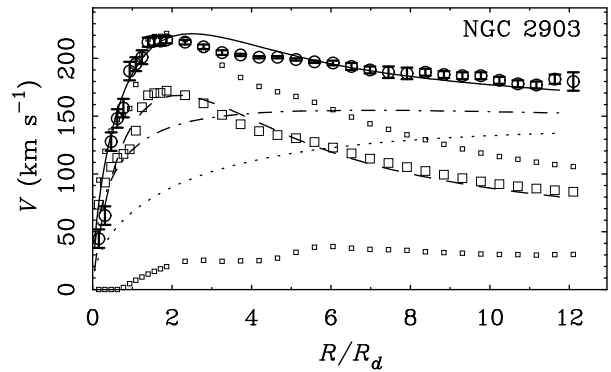,width=.9\hsize,clip=}}
\caption{The rotation curve and mass model of the luminous high
surface brightness galaxy NGC 2903.  The rotation curve data are
plotted as circles with error bars.  The contribution to the rotation
by the baryonic component (stars plus gas) is denoted by the square
symbols.  Large squares are for the modeled mass-to-light ratio
specified in Table \ref{datafits}.  Also shown as small squares are
the limiting cases of minimum (gas only with $\ML = 0$) and maximum
disk.  The dashed line shows the adiabatically formed exponential disk
used to approximate the observed baryon distribution.  The solid line
is the total rotation due to disk plus compressed halo.  The
primordial NFW halo (with parameters given in Table \ref{datafits}) is
shown by the dotted line, and the compressed halo by the dash-dotted
line.
\label{NGC2903}}
\end{figure}

The choice of stellar mass-to-light ratio has a pronounced effect on
the mass model.  There is a tremendous range of possibilities between
maximum and minimum disk (e.g, Fig.~\ref{NGC2903}).  Attempts to grow
a maximum disk in an NFW halo inevitably lead to rotation curves which
peak too high and fall too far before flattening out.  For this
reason, maximum disks and halos with central cusps appear to be
mutually exclusive.

\begin{figure}[t]
\centerline{\psfig{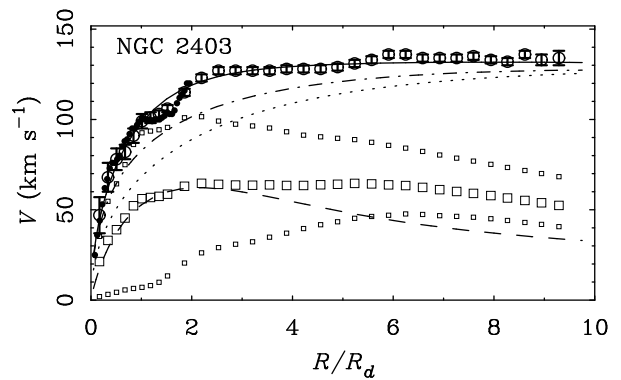}}
\caption{The intermediate luminosity, high surface brightness galaxy
NGC 2043.  All symbols and lines are the same as for
Fig.~\ref{NGC2903}.  In addition to the extended HI rotation curve of
Begeman (1987), the Fabry-Perot data of Blais-Ouellette \etal\ (2004)
are shown as the small solid points in the rising part of the rotation
curve.  The extended HI in this galaxy makes it difficult to
approximate the baryons with a single exponential disk.
\label{NGC2403}}
\end{figure}

Nevertheless, a heavy disk is important for falling rotation curves.
The disk fraction \md\ modulates the shapes of rotation curves.
Systematic variation of \md\ with rotation velocity appears to be
required in order to match the observed trend in rotation curve shapes
from falling for bright galaxies to rising for dwarfs (e.g., Persic \&
Salucci 1991).  This variation can be rapid, with $\md \approx 0.1$
for NGC 2903 falling to $\md \approx 0.02$ for NGC 2403
(Fig.~\ref{NGC2403}).

Systematic variation of \md\ has important ramifications for the
Tully-Fisher relation.  It implies that the observed range of baryonic
disks are all found in a relatively narrow range of halo masses.  Such
a finding changes the slope of the Tully-Fisher relation from that
nominally predicted for halos only.

\begin{figure}[t]
\centerline{\psfig{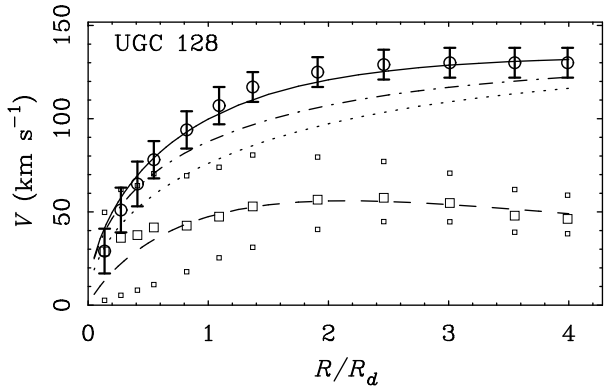}}
\caption{The intermediate luminosity, low surface brightness galaxy
UGC 128.  All symbols and lines are the same as for
Fig.~\ref{NGC2903}.  This galaxy is well matched to NGC 2403 in terms
of luminosity and circular velocity: these two objects occupy the same
location in the Tully-Fisher diagram (de Blok \& McGaugh 1996).  An
approximation to the baryonic component is obtained by stretching the
exponential disk scale length to 10 kpc, somewhat longer than that of
the stars, which varies from $\sim 6$ to 9 kpc, depending on bandpass
(de Blok \etal\ 1995).
\label{UGC128}}
\end{figure}

UGC 128 (Fig. \ref{UGC128}) is comparable to NGC 2403 in luminosity
and rotation velocity.  Despite the low surface brightness of this
galaxy, the halo is perceptibly compressed: the disk is not completely
negligible.  The disk fraction and \ld\ are very similar to those of
NGC 2403.  This is hardly surprising, since the rotation curves of the
two galaxies are indistinguishable when normalized by scale length:
$V(r)$ is very different, but $V(r/\rd)$ is much the same (de Blok \&
McGaugh 1996).  Consequently, the density of the halo of UGC 128 must
be lower than that of NGC 2403, simply because \rs\ is larger at
comparable mass.  This is reflected in a low concentration ($c \approx
3$).  Though the fit with a compressed NFW halo is good, primordial
halo concentrations this low never occur in any plausible cosmology
(Navarro \etal\ 1997; McGaugh \etal\ 2003).

\begin{figure}[t]
\centerline{\psfig{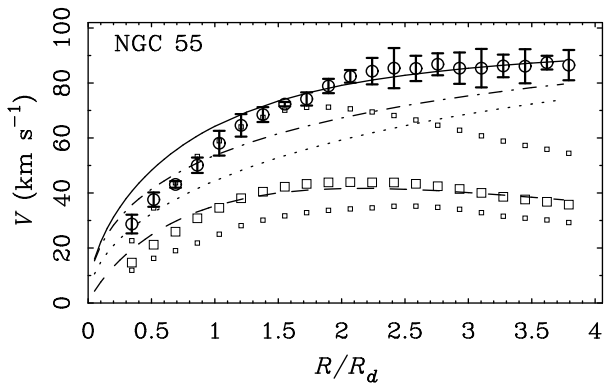}}
\caption{NGC 55 is a high surface brightness dwarf galaxy.  All
symbols and lines are the same as for Fig. \ref{NGC2903}.  The
baryonic rotation curve is approximated by an exponential disk of
scale length 2.7 kpc (vs.\ 1.6 kpc for the stars alone).
\label{NGC55}}
\end{figure}

\begin{figure}[t]
\centerline{\psfig{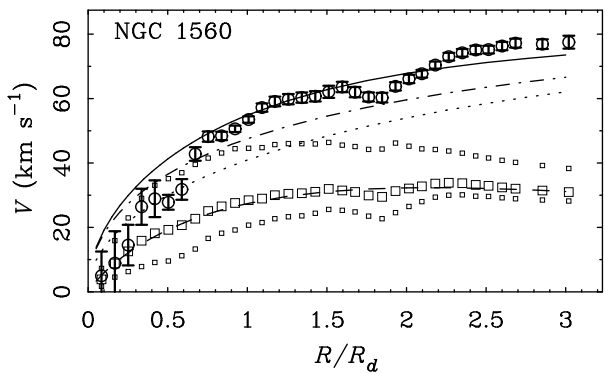}}
\caption{NGC 1560 is a low surface brightness dwarf galaxy.  All
symbols and lines are the same as for Fig. \ref{NGC2903}.  The
baryonic rotation curve is approximated by an exponential disk of
scale length 2.6 kpc (vs.\ 1.3 kpc for the stars alone).
\label{NGC1560}}
\end{figure}

NGC 55 (Fig.~\ref{NGC55}) and NGC 1560 (Fig.~\ref{NGC1560}) are nearby
dwarf galaxies.  The gas is not negligible in these galaxies.  The
baryonic rotation curve is reasonably well approximated by a single
exponential with a scale length stretched relative to that of the
stars alone (roughly by a factor of two) to represent both stars and
gas.

In order to produce something like the observed, gradually rising
rotation curve, it is necessary to consider very small disk fractions.
The cases illustrated with $\md = 0.01$ provide a good fit to the
baryonic components, and a reasonable match to the outer parts of the
rotation curves.  The fit to the inner few kpc of the rotation curves
is not good.  The problem here stems from the assumed form of the
primordial halo.  The NFW halo is too centrally concentrated to begin
with, a well known problem for dwarf galaxies (C\^ot\'e, Carignan \&
Freeman 2000; de Blok \etal\ 2001; Swaters \etal\ 2003; de Blok, Bosma
\& McGaugh 2003).

\section{Conclusions}
We have used the algorithm described by Young (1980), which takes
proper account of radial motion, to compute the density of a dark
matter halo that is compressed by the growth of a disk in its center.
Young developed the method for compression of a spherical star cluster
by growth of a central black hole; the algorithm was applied
independently for halo compression by Wilson (2004).  While Young's
algorithm is more complicated to program, and more time consuming to
run, than the simple and popular algorithm (Blumenthal \etal\ 1986),
it correctly predicts rather less halo compression, especially in the
inner region that is most accessible to observations.

The predicted compressed density profiles agree almost perfectly with
those found in high-quality $N$-body simulations
(Fig.~\ref{sphcheck}), and are significantly less compressed than
predicted by the Blumenthal algorithm.  The almost perfect agreement
with $N$-body simulations confirms also that the spherical
approximation for the disk mass profile is entirely adequate, as
previously found by Barnes (1987), Jesseit \etal\ (2002), and Wilson
(2004).

The reason that halos are less easily compressed than the Blumenthal
algorithm would predict is that this crude method neglects the extra
pressure caused by radial motions.  We present tests
(Fig.~\ref{anisot}) to show that the compressed density is lowered as
the importance of radial motion is increased.  Since dark matter halos
are formed in a hierarchical collapse, the outer halos are likely to
have radially biased velocity distributions, although the inner halos
are expected to be close to isotropic.

The algorithm can be applied to all spherical density profiles having
valid distribution functions, and any arbitrary mass profile for the
disk.  The key assumptions, that the halo be spherical and that the
disk mass was assembled adiabatically, are no more restrictive than
for the usual Blumenthal algorithm.  It is expected that hierarchical
galaxy formation leads to aspherical dark matter halos, and
compression and shape changes can be computed in these more realistic
circumstances only by $N$-body simulation (\eg\ Kazantzidis \etal\
2005).

The adiabatic assumption is likely to hold if the disk remains closely
axisymmetric, as the cooling and settling of gas is likely to occur
over many crossing times in the inner halo.  However, non-adiabatic
resonant interactions will develop between rotating, non-axisymmetric
structures in the disk and the orbits of halo particles (Tremaine \&
Weinberg 1984).  It is unclear at present whether such interactions
can cause significant changes to the halo density profile (Weinberg \&
Katz 2002; Sellwood 2003).

As proof of principle only, we have derived the parameters of initial,
uncompressed NFW halos for a few well-studied galaxies, and plan to
apply the technique to a larger, and more representative, sample of
galaxies in a later paper.  The initial results are promising in many
respects.  We found it easy to obtain reasonable fits, with variations
in disk fraction naturally explaining the variation in shapes of
rotation curves.  However, it does not appear likely that an improved
treatment of halo compression will provide a resolution of the
cusp/core controversy, or of the absolute halo density of low surface
brightness galaxies, which remains a problem for \LCDM.

\acknowledgments We thank S. Sridhar for drawing our attention to
Young's paper and Greg Wilson and Agris Kalnajs for informing us of
their work in advance of publication.  The referee, Oleg Gnedin,
provided a helpful report.  This work was supported by grants
AST-0507323 and NNG05GC29G to JAS and AST-0206078 and NAG513108 to
SSM.

\end{document}